\documentclass{elsart}

\usepackage{graphicx}
\usepackage{dcolumn}
\usepackage{bm}
\usepackage{amssymb}

\def \ee{\end{equation}}
\def \be{\begin{equation}}

\begin{document}
\begin{frontmatter}

\title{Exclusion of black hole disaster scenarios at the LHC}

\author{Benjamin Koch}

\address{
Pontificia Universidad Cat\'{o}lica de Chile, \\
Av. Vicu\~{n}a Mackenna 4860, \\
Santiago, Chile \\
\& \\
Institut f\"ur Theoretische Physik, Johann Wolfgang Goethe -
Universit\"{a}t,\\
Max-von-Laue-Strasse 1\\
D-60438 Frankfurt am Main, Germany 
}
\ead{bkoch@fis.puc.cl}

\author{Marcus Bleicher}
\address{
Institut f\"ur Theoretische Physik, Johann Wolfgang Goethe -
Universit\"{a}t,\\
Max-von-Laue-Strasse 1,\\
D-60438 Frankfurt am Main, Germany}

\author{Horst St\"{o}cker}

\address{
Institut f\"ur Theoretische Physik, Johann Wolfgang Goethe -
Universit\"{a}t,\\
Max-von-Laue-Strasse 1,\\
D-60438 Frankfurt am Main, Germany \\
\& \\
Frankfurt Institute for Advanced Studies (FIAS),\\
Ruth-Moufang-Strasse 1, \\
D-60438 Frankfurt am Main, Germany\\
\& \\
GSI - Helmholtzzentrum f\"ur Schwerionenforschung GmbH,\\
Planckstrasse 1,\\
D-64291 Darmstadt, Germany\\
}

\date{\today}

\begin{abstract}
The upcoming high energy experiments at the LHC
are one of the most outstanding efforts for a better understanding
of nature. It is associated with great hopes in the physics community.
But there is also some fear in the public, that the conjectured
production of mini black holes might lead to a dangerous chain reaction.
In this paper we summarize the most straight forward arguments
that are necessary to rule out such doomsday scenarios.
\end{abstract}
\end{frontmatter}

\section{Motivation}

As an explanation for the large hierarchy between the
Planck scale and
the electroweak scale some authors
 postulated the existence
of additional spatial dimensions
\cite{ArkaniHamed:1998rs,Antoniadis:1998ig,ArkaniHamed:1998nn,Randall:1999ee,Randall:1999vf}.
One exciting consequence of such theories is that they allow for the production of black
holes in high energetic particle collisions
\cite{Banks:1999gd,Giddings:2000ay,Giddings:2001bu,Dimopoulos:2001hw,Hossenfelder:2001dn,Bleicher:2001kh,Kotwal:2002wg,Giddings:2004xy}.
It was further conjectured that black holes could have a stable final state. This lead to a
public discussion whether such mini black holes once they are produced at the large hadron
collider (LHC) could be growing dangerously inside the earth \cite{blogs}. 
There is to our knowledge no scientific work that
predicts that the remnants (if they exist) of such mini black holes (if they exist)
could be stable at masses far above the Planck scale $M_f$.
However, given the public alarm over the subject, we 
want to go further and also exclude danger from scenarios which are to
the present understanding of the physics of mini black holes not well motivated.
A number of counter arguments disfavor such disaster scenarios. Recently those arguments
have been summarized and discussed by a group \cite{Giddings:2008gr} who comes to
the conclusion that ``there is no risk of any significance whatsoever from such black holes".
In this paper we independently present a short coherent argument why there is no risk due to
mini black holes from TeV particle collisions. First we look at the logically possible black
hole evolution paths. After this we show for every endpoint of the paths, why mini black
holes can not be dangerously growing. For this we use arguments which 
are already present in
\cite{Giddings:2008gr}, but we also bring forward new arguments such as 
the influence of a strongly growing
black hole mass on the escape velocity of the mini black hole.

\section{Black holes in large extra dimensions}
High energy experiments like those at the large hadron collider (LHC) play a crucial role for
a better understanding of the fundamental laws of physics. One hope is that those experiments
can discriminate between several approaches that try to extend the physical framework of the
standard model
\cite{Dimopoulos:2001hw,Bachacou:1999zb,Khoze:2001xm,Dimopoulos:2001qe,Weiglein:2004hn,ArkaniHamed:2004fb,Lillie:2007yh,Cheung:2007zza}.
In some models
\cite{ArkaniHamed:1998rs,Antoniadis:1998ig,ArkaniHamed:1998nn,Randall:1999ee,Randall:1999vf}
it was conjectured that the hierarchy problem between the Planck scale, $m_{Planck}\approx
10^{19}$~GeV, and the electroweak scale, $m_{EW}\approx 100$~GeV, can be solved
 by postulating the existence of additional spatial dimensions.
In reference  \cite{ArkaniHamed:1998rs,Antoniadis:1998ig,ArkaniHamed:1998nn} this is done by assuming that the (d)
additional spatial dimensions are compactified on a small radius $R$ and
further demanding that all known Standard Model particles exist on a $3+1$ dimensional
sub-manifold ($3-$brane). They find that the fundamental mass $M_f$ and the Planck mass
$m_{Plank}$ are related by
\begin{eqnarray}
m_{Planck}^2 = M_f^{d+2} R^d \quad. \label{Master}
\end{eqnarray}
Within this approach it is possible to have a fundamental gravitational scale of  $M_f\sim
1$~TeV. The huge hierarchy between $m_{EW}$ and $m_{Planck}$ would then come as a result of
our ''ignorance'' regarding extra spatial dimensions. Due to the comparatively low
fundamental scale $M_f\sim$~TeV and the hoop conjecture \cite{hoop}, it might be possible to
produce mini black holes with masses of approximately 1 TeV in future colliders
\cite{Banks:1999gd,Giddings:2000ay,Giddings:2001bu,Dimopoulos:2001hw,Hossenfelder:2001dn,Bleicher:2001kh,Kotwal:2002wg,Giddings:2004xy}.
This can only be the case when the invariant scattering energy $\sqrt{s}$ reaches the
relevant energy scale $M_f$. The higher dimensional Schwarzschild radius
\cite{Myers:1986un,Giddings:2001bu} of these black holes is given by
\begin{equation} \label{ssradD}
R_H^{d+1}=
\frac{16 \pi (2 \pi)^d}{(d+2) A_{d+2}}\left(\frac{1}{M_{f}}\right)^{d+1} \; \frac{M}{M_{f}}
\quad ,
\end{equation}
where $A_{d+2}$ is the area of the $d+2$ dimensional unit sphere 
\be A_{d+2}=\frac{2
\pi^{(3+d)/2}}{\Gamma\left(\frac{3+d}{2}\right)}\quad. 
\ee 
A semi-classical approximation for the
mini black hole production cross section is given by
\begin{eqnarray} \label{cross}
\sigma(M)\approx \pi R_H^2 \xi(\sqrt{s}-M_f)\quad,
\end{eqnarray}
where the function $\xi$ ensures that black holes are only produced above the $M_f$
threshold. The function $\xi$ is one for $\sqrt{s}\gg M_f$ and zero for $\sqrt{s}\approx
M_f$. In many simulations $\xi$ is replaced by a theta function. The validity of this
approximation has been debated in
\cite{Voloshin:2001fe,Voloshin:2001vs,Giddings:2001ih,Rizzo:2001dk,Jevicki:2002fq,Eardley:2002re,Rychkov:2004sf,Rychkov:2004tw,Kang:2004yk,Rizzo:2006uz,Rizzo:2006zb}
and the observable formation of an event horizon has been questioned
\cite{Vachaspati:2006ki,Vachaspati:2007hr}. 
However, other improved calculations including
the diffuseness of the scattering particles (as opposed to point particles) and the angular
momentum  of the collision (as opposed to head on collisions) as well as string inspired
arguments only lead to modifications of (\ref{cross}) which are of the order of one
\cite{Yoshino:2002tx,Solodukhin:2002ui,Ida:2002ez,Horowitz:2002mw}. This would open up a
unique possibility of studying gravitational effects at very small distance scales in the
laboratory. Such observations of gravitational physics at the tiny scales of the quantum
world may provide access to the presently biggest question of theoretical physics: A unified
description of quantum physics and gravity.

At the same time there is a growing concern in the public. ``Could such monstrous objects like
mini black holes (once they are produced at LHC) eat up the entire world?" This question is
controversially discussed in blogs and online-video-portals \cite{blogs}. Similar anxieties (with
strangelets instead of black holes) have already been stirred up when the previous generation
of collider was built \cite{Dar:1999ac}. 
Fears of possibly dangerous mini black holes have been augmented by
the idea of a quasi stable black hole final state.
A quasi stable black hole final state has
been frequently studied in the literature
 \cite{Whitt:1985ki,Aharonov:1987tp,Gibbons:1987ps,Whitt:1988ax,Bowick:1988xh,Callan:1988hs,Myers:1988ze,Coleman:1991sj,Lee:1991qs,Banks:1992ba,Barrow:1992hq,Banks:1992is,Maeda:1993ap,Giddings:1993vj,Bonanno:2000ep,Adler:2001vs,Alexeyev:2002tg,Baker:2003ds,Hossenfelder:2003dy,Koch:2005ks,Hossenfelder:2005ku,Humanic:2006xg,Ghaffarnejad:2007tm,Li:2007ga,Koch:2007yt,Nicolini:2008aj} which
partially refer to astrophysical black holes and partially refer to mini black holes. 
Instead of ignoring this concern we take it serious
and try to discuss the issue without provoking an emotional palaver. We explain from
theoretical arguments why such a disaster is generally believed to be impossible. But we even
go one step further and discuss the question: ``What if the theory is wrong?" We show that
even if the current theories are wrong, there is no danger as long as the ``true theory" is
not completely unphysical \cite{Arkani:2008}.
By mostly using arguments that are based on
black hole production in high energetic cosmic rays \cite{Ringwald:2001vk},
a recent and extensive study on the (im) possibility of dangerous
mini black holes has been given in \cite{Giddings:2008gr}.
However, in this paper we want to concentrate
on a short but convincing argument.

\section{Possible black hole evolution paths}

\begin{figure}[h]
\includegraphics[width=14cm]{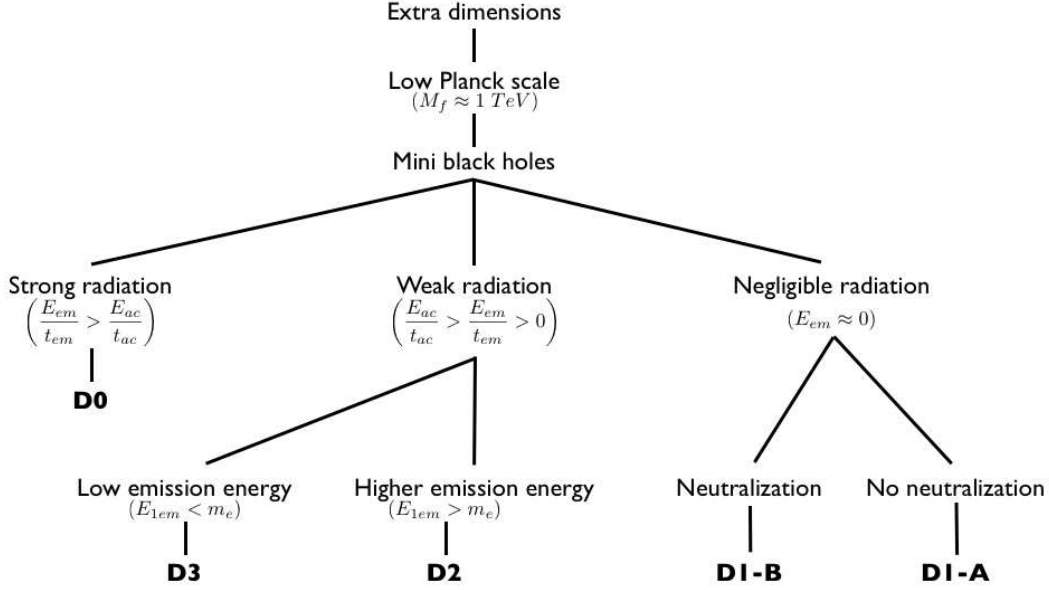}
\caption{\label{BHtree}Possible black hole evolution paths.}
\end{figure}

The logical structure of the assumptions that are relevant for this study is shown in figure
(\ref{BHtree}). We will now discuss the tree structure in figure (\ref{BHtree}) step by step.
Every branch of the tree ends with a discussion ({\bf{D0-D3}}) which can be found 
in the next section.
In those discussions we explain with either theoretical or experimental
arguments why the discussed branch can not have any disastrous consequences.
Therefore, we define the average energy $(E_{em})$ as the energy which is emitted
in the rest frame of the mini black hole
in the average time scale $(t_{em})$.
The corresponding definition for accretion gives 
the average energy $(E_{ac})$ as the energy which is accreted
in the rest frame of the mini black hole
in the average time scale $(t_{ac})$.
If not explicitly stated otherwise, accretion times and energies
are those for relativistic mini black holes from high energetic cosmic rays.
In order to open up the possibility of producing mini black holes in a $14$~TeV collider, one
has to assume the existence of extra dimensions with a fundamental mass scale in the
$\sim$TeV range. Next one has to assume that quantum gravity effects do not spoil the
conjecture that classical closed trapped surfaces lead to the formation of a black hole event
horizon. If all this is given then the mini black hole could in principle follow three
different paths in its further development.
First, it could emit high energetic radiation
($E_{em}$) in a short time scale ($t_{em}$) such 
that a comparison to the accretion energy
($E_{ac}$) and accretion time
 ($t_{ac}$) shows a net emission
 ($E_{em}/t_{em}>E_{ac}/t_{ac}$).
This is what most theoreticians predict and it would be the case for both, the balding phase and the Hawking phase. In the tree (\ref{BHtree}) this possibility is denoted as
''{\it{Strong radiation}}''. As discussed in ({\bf{D0}}) such a black hole can not cause any
danger.

Secondly, the mini black hole might (in contradiction
to Hawking's calculation) not emit any radiation which is caused
by the curvature of space time. 
In the tree this possibility is denoted as ''{\it{Negligible radiation}}''.
In this case it consumes
everything it encounters on its trajectory. By this it should acquire some net charge. 
As discussed in ({\bf{D1-A}}) 
mini black holes with this property are ruled out by high energetic cosmic ray observations.
One could further assume that the acquired net charge is radiated away without
loosing a significant amount of energy. This case is discussed in ({\bf{D1-B}})
for two complementary scenarios which both 
show that high energy cosmic ray observations
rule out any danger from such mini black holes.

The third possibility is a relatively weak radiating black hole (eventually forming a
black hole remnant). This means that the mini black hole eats in average more matter than it emits $E_{ac}/t_{ac}>E_{em}/t_{em}>0$, it is 
therefore labeled by "{\it{Weak radiation}}". 
In this case one can distinguish between the two cases
where the emission energy per particle $(E_{1em})$ in the rest frame of the mini
black hole is either larger or smaller than the electron mass $(m_e)$.
As shown in the discussions  ({\bf{D2}}) and  ({\bf{D3}}),
both cases inevitably lead to a stopping of mini black holes
from cosmic rays which rules out any danger from the concerning
scenarios.

Please note that the structure of the different evolution paths
is held in such a way that it can not be mistrusted by arguments that
refer to a possibly different radiation mechanism. For instance also
a conjectured neutralization of the black holes via a Schwinger mechanism 
is covered by discussion ({\bf{D1-B}}).

\section{Discussions}

\subsection*{{\bf{D0}}: the black hole temperature}
Most theoretical models for large extra dimensions predict that the mini black holes emit
high energetic radiation in a very short time scale. The temperature of this radiation was
derived from the quantum theory in curved spacetime \cite{Hawking:1974sw,Hawking:1982dh}.
This so called Hawking temperature is inverse proportional to the radius of the black hole
\cite{Giddings:2001bu} 
\be T_H=\frac{d+1}{4 \pi R_H}\quad. 
\ee 
The time scale of a single
emission can be straight forward estimated $t_{1em}\approx  R_H/c $. Therefore, the decay rate
of a mini black hole in the canonical picture is 
\be\label{BHdecay} \frac{dM}{dt}\approx- c
\frac{d+1}{4 R_H^2} \quad. \ee 
Comparing this decay rate to typical growth rates
at the early stage of the mini black hole evolution
\cite{Giddings:2008gr} one finds for instance 
for $M=10 M_f = 10^4$GeV that the decay rate
(\ref{BHdecay}) exceeds the growth rate for any number of dimensions by at
least thirty orders of magnitude. From this estimate it is clear that such 
mini black holes that are produced on the earth can never grow.

\subsection*{{\bf{D1-A}}: the black hole charge}
In this discussion we rely on the logical imperative that those mini black holes that
originate from the collapse 
of charged particles or that swallow charged particles also have
effectively some charge.
If the average time for the emission of a single particle in the
rest frame of the black hole $(t_{1em})$ is of the same order of magnitude
or even bigger than
the average time ($t_{ac}$) between accretion events ($t_{1em}\gtrsim t_{ac}$)
one can make a simple
random walk approximation for the black hole charge.
In this approximation, 
the charge that a black hole inherits from the accreted or collapsed matter
is 
\be |Q_e|=\frac{\sum_i |q_i | t_i}{\sum_k
t_k}\quad, 
\ee 
where $q_i$ is the charge of the black hole at each step of its evolution and
$t_i$ is the duration of this step of the evolution. 
If the black hole charge changes
randomly at any step of the evolution it scales like ($\sqrt{n}$) where ($n$) is the number of
steps. But even in the unlikely case that the black hole always tends to neutralize in the
following step after obtaining the charge $|q_i|\ge 1/3$ (for a quark), this still results in
an effective charge $|Q_e|\ge 0.16$, if the neutralization time scale
is at least of the same order of magnitude as the accretion time scale. 
If such black holes would be produced at the highest
center of mass energy at the LHC ($\sqrt s_{NN}=14$ TeV), then black holes must have been
produced in the whole past life time of the earth and the sun from high energetic cosmic ray
events (having an even higher center of mass energy of up to $\sqrt s_{NN}=400$~TeV).
However, there is
one difference between the black holes from cosmic rays and those in the laboratory: The
black holes in the laboratory might have a very low kinetic energy (i.e. velocity) in the
rest frame of the earth, while the black holes from cosmic rays always have at least a
momentum of 
\be p\ge \frac{M^2}{2 m_p}\left(1\pm \mathcal{O}\left( \frac{m_p}{M}\right)
\right)\quad, 
\ee 
where $m_p$ is the mass of a proton. This means that a black hole from a
cosmic ray with a rest-mass of $\sim1$~TeV has a kinetic energy of at least $\sim0.5\;
10^6$~GeV. The kinetic energy loss of a black hole with an effective electrical charge $|Q_e|$ and a
mass $M$ can be calculated with the help of the Bethe-Bloch formula 
\be\label{BeBlo}
\frac{dE}{dx}= \frac{\kappa}{A_0}\frac{|Q_e|^2 Z}{\beta^2} \left(\frac{1}{2} \log \left(2 m_e
\frac{\beta^2 T_{m}}{I^2 (1-\beta^2)}\right)\right) \quad, 
\ee 
where
($\beta=\sqrt{1-M^2/(M^2+E_{kin}^2)}$),  
($Z$) is the average charge of the target,
($T_m=2 m_e \beta^2/(1-\beta^2) 1/(1+2 m_e/(M
\sqrt{1-\beta^2})+m_e^2/M^2)$),  ($I$) is the average
electronic excitation levels of the target, ($m_e$) is the electron mass, and ($\kappa/A_0$) is
the standard energy loss parameter of the target.
The resulting curves from equation (\ref{BeBlo}) have a minimum at relatively low kinetic energies
and a logarithmic growth for higher energies. It is also clear that a higher effective charge
also means a higher energy loss since the energy loss is ($\sim |Q_e|^2$). By only taking the
minimum energy loss of those curves one finds that ($\sim 1$~TeV) black holes can be stopped in
the earth if they effectively carry ($Q_e > 0.4$) elementary charges. The whole argument can be
extended by replacing the earth by the sun which shows that mini black holes from 
cosmic rays with an
effective charge of ($Q_e > 0.04$) would be stopped in the sun. 
Since the expected effective charge is ($|Q_e|\ge 0.16$), we
can conclude that the existence of our solar system proofs that mini black holes can not be
dangerous because they would have already been produced and stopped inside the
earth (sun) without causing any damage. Although this argument is sufficient to rule out
dangerous charged black holes we want to mention that it underestimates the true stopping
power by far. Especially taking into account the dense core of the sun and the process of
pair creation in the Bethe-Bloch formula increases the effect by at least three orders of
magnitude \cite{Giddings:2008gr}.

\subsection*{{\bf{D1-B}}: neutralization without significant energy loss}

In this discussion we refer to scenario where it is assumed that
black hole radiation is only present if the black hole is charged.
This very special scenario of black hole evolution has been put forward
by the assumption that a black hole only radiates if it is charged  
\cite{plaga,Plaga:2008gq}
and the Hawking radiation is strongly suppressed.
Such a behavior was motivated by postulating a Schwinger mechanism
and a coincidental suppression of Hawking or Unruh radiation.
This scenario seems to be especially tuned to make
the microscopical black holes grow without being slowed down by
electromagnetic interactions. 
However, as explained in  \cite{Giddings:2008gr} 
such a scenario is highly doubtable. 
The reason is that there is no known mechanism 
to shut off the quantum effects responsible for Hawking radiation, but still leave intact 
either the quantum effects responsible for Schwinger discharge, or some other neutralization
mechanism that acts to discharge the black holes. 
Since the time scale of such a neutralization due to a Schwinger process
is supposably extremely short ($\sim R_H/c$), the discussion ({\bf{D1-B}})
of an effectively charged black hole that gets slowed down by electromagnetic
interactions can not be applied. 
Therefore, the only possible
process to slow down such a mini black hole is the accretion slow down.
We explicitly consider two straight forward equations to describe the growth
of the black hole (which originates from a high energetic
cosmic ray collision) propagating through an aggregation of matter.
Both equations can be seen as complementary simplifications of
a realistic description. We applying those growth equations and
their effect due to accretion slow down to different 
astronomical objects and find that
cosmic ray arguments also exclude any danger from those mini black holes.

For the first equation it is assumed 
that the accretion process for a mini black hole
who has some overlap with the wave function of a nucleon is dominated by
the strong interaction. 
Remember that the nucleon radius $r_p$ is much 
larger than the black hole radius $r_H$. 
As soon as some colored part
of the nucleon is trapped inside of the black hole, all subsequent
dynamics could be dictated by the strong interaction between the 
remaining nucleon color charges and the trapped color charges.
Thus, in this first approximation we assume that the possible (color neutral)
final states of a black hole - nucleon system after such an encounter 
are only determined by strong dynamics while the effects of
the black hole rapidity and of the black hole surface are neglected.
Those strongly interacting but color neutral final states 
(with a neutralized Black hole $BH$)
for an initial black hole - proton system could be 
($BH + p^+ \rightarrow \{BH+ e^+, BH + \pi^+ , BH +  \pi^+ + \pi^0 , \dots, BH + X  \}$).
In order to be most pessimistic about the braking efficiency of the reaction
one has to assume that no momentum is transferred to the final state ($X$).
In this case the mini black hole grows in mass if the rest mass of 
($X$) is smaller than the rest mass of the nucleon. 
This effect of the strong dynamics can be parameterized by claiming
that a black hole can accrete some fraction ($1>\alpha>0$) of a nucleon
when it propagates through it. 
For this kind of accretion the mass growth ($dM$) after propagating a
distance ($dx$) in a star with average density ($\rho$) is at least
\be\label{Mgrowth}
\frac{dM_1(x)}{dx}=\pi (r_p+R_H)^2 \rho \alpha
\ge \alpha \pi r_p^2 \rho \quad
\Rightarrow \quad M_1(x)\ge x  \pi r_p^2 \rho \alpha+M_f \;.
\ee
This solution is independent of the number of extra dimensions $d$ and valid 
within its assumptions
as long as the black hole radius is much smaller than the nucleon radius 
(for $d\ge1$ this means $M_1<5\times10^9$~GeV).
The subscript in ($M_1$) refers to the fact that this is evolution scenario number one.
The approximation (\ref{Mgrowth}) is tuned to represent the subsequent
accretion of a nucleus, but it neglects a possible rapidity and area dependence
of the black hole accretion. 
In this sense it can be seen as complementary to the following more standard
accretion estimate, which neglects the strong dynamics but takes
into account the geometric area of the black hole.

The second equation is more intuitive than (\ref{Mgrowth})
and it is also the basis of the discussion in \cite{Giddings:2008gr}.
Here it is assumed that the
black hole consumes and keeps all matter and energy that passes its
trajectory. 
In this case one can estimate 
the black hole growth rate to be proportional to the black holes surface area
\be\label{area} A(M)=4 \pi \left(\frac{16 \pi (2 \pi)^d}{(d+2) A_{d+2}}\right)^{2/(d+1)}
\frac{1}{M_f^2}\left(\frac{M}{M_f}\right)^{2/(d+1)}\quad. 
\ee 
With the area (\ref{area}) and
the density of the matter ($\rho$), the growth rate is at least
\be\label{Mgrowth2}
\frac{dM_2(x)}{dx}= \rho A(M_2)\quad. 
\ee
The subscript in ($M_2$) refers to the fact that this is evolution scenario number two.
Equation (\ref{Mgrowth2}) suggests that
for ($d>1$) the growth rate due to thermal motion inside
of the earth would be too slow to do any harm within the lifetime of the earth.
However, for low black hole velocities (as they would be possible at LHC)
equation (\ref{Mgrowth2}) has to be corrected by taking additional effects
such as Bondi accretion into account. It turns out that only
for $d>6$ the growth rate due to thermal motion inside
of the earth would be too slow to do any harm within the lifetime of the earth  
\cite{Giddings:2008gr}.
Therefore, we have to study the cases $d\le 6$ of (\ref{Mgrowth2}).
Also this second estimate has its weakness because it does
not take into account any effects due to the strong interaction inside a nucleon.
A further weakness is that it does not
respect the consistency condition due microgravity experiments
that confirm a $(3+1)$ dimensional behavior down to the micrometer scale,
but it stays pretty robust for high black hole rapidity.
Even though the growth equations (\ref{Mgrowth}, \ref{Mgrowth2}) only apply
in certain limits, we use them as complementary ends of 
more elaborate approximations  \cite{Giddings:2008gr}.

\begin{figure}[h!]
\includegraphics[width=12cm]{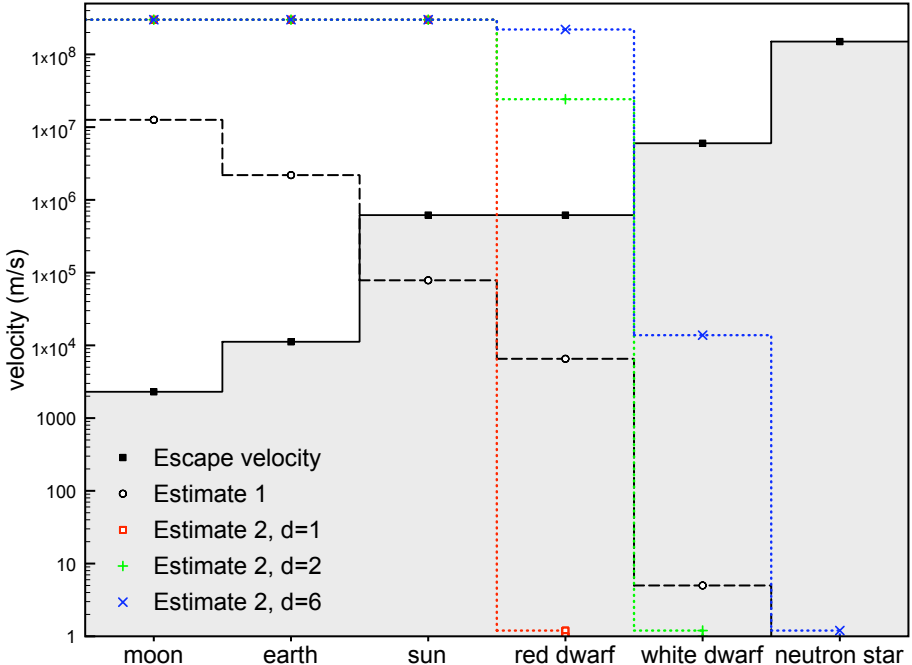}
\caption{\label{velocities}
Escape velocity compared to the
velocity of an originally TeV black hole after propagating
through the moon, the earth, the sun, a red dwarf, a white
dwarf, or a neutron star 
according to  equations (\ref{Mgrowth})$\equiv$Estimate 1, (\ref{Mgrowth2})$\equiv$Estimate 2.
In equation (\ref{Mgrowth}) $\alpha$ was set to one.
For equation (\ref{Mgrowth2}) the cases $d=\{1,2,6\}$ are plotted,
the remaining cases $d=\{3,4,5\}$ lie between those curves.
For the radii and densities average values were taken.
}
\end{figure}

Now we apply the growth equations (\ref{Mgrowth}, \ref{Mgrowth2})
to the accretion slow down, which is the only mechanism of slowing
down the neutralizing mini black holes of this discussion after they are produced in cosmic rays.
As mentioned in \cite{Giddings:2008gr}, the accretion slow down
does in the worst case of ``perfect accretion'' not lead to any momentum
transfer. However, one can use the relativistic relation between the black hole rest
mass $M_i$, energy $E_i$, and velocity $v_i$
\be
E_i=\frac{M_i}{\sqrt{1-\frac{v_i^2}{c^2}}}\quad,
\ee
where the index ($i=1$) refers to scenario (\ref{Mgrowth})
and the index ($i=2$) refers to scenario (\ref{Mgrowth2}).
Now one can solve this equation for the velocity ($v_i$) and
use the relativistic energy-momentum relation ($E_i^2=M_i^2+p^2$) which
gives 
\be
v_i\approx c\sqrt{1-\frac{M^2_i}{M_i^2+p^2}}\quad.
\ee
The momentum ($p$) of the mini black hole in the case of  ``perfect accretion''
does not change during the accretion 
and is therefore the momentum it inherits from its production
due to an high energetic cosmic ray event. Close
to the production threshold of ($M_f\approx 1 $~TeV) this momentum as seen from
the laboratory frame is ($p \approx M_f^2/(2 m_p)$), where ($m_p$) is the proton mass.
Thus, the velocity of the mini black hole after
propagating through a star (planet) with radius ($r$) reads
\be\label{myvel}
v_i\approx c\sqrt{1-\frac{M^2_i(2r)}{M_i^2(2r)+M_f^4/(2 m_p)^2}}\quad.
\ee
This equation shows that even in the case of ``perfect accretion''  without momentum
transfer, the mini black hole velocity can decrease due to its mass growth
which was not take into account by \cite{Giddings:2008gr}.
Please note that this decrease of the black hole velocity is solely an effect of the growing
black hole mass while other speed diminishing interactions 
(that in most scenarios play a dominant role) are not even taken into account.
Those velocities are compared to according escape velocities in figure (\ref{velocities}).
As can bee seen in figure (\ref{velocities}), equation (\ref{Mgrowth}) already 
leads to a contradiction with the existence of stars like the sun and their 
corresponding escape velocities.
The next exclusion comes for equation  (\ref{Mgrowth2}) with $(d=1)$ 
and the existence of red dwarfs.
Figure  (\ref{velocities}) further shows that 
for equation (\ref{Mgrowth2}) with $(6 \ge d>1)$
there is a contradiction to the existence of white dwarfs and the corresponding escape velocities.
Thus, one can say that the existence of old ($> 1 gyr$) white dwarfs
is in contradiction to dangerous black holes that behave according to equations 
(\ref{Mgrowth} or \ref{Mgrowth2}).
The result would be even clearer for neutron stars but
not all neutron stars can be
used as an argument because the ultra high energetic cosmic rays undergo deflection and
deceleration in the large magnetic fields that exist around neutron stars \cite{Giddings:2008gr}.
For equation (\ref{Mgrowth}) the argument with the white dwarfs works as long
as the fraction $\alpha$ is bigger than $1 \times 10^{-7}$.
This limit is obtained by varying $\alpha$ in equations (\ref{Mgrowth},\ref{myvel})
and comparing the result to the escape velocity of a white dwarf.
But what happens when the fraction is smaller than $1 \times10^{-7}$?
In this case one can look at the growth rate of a mini black hole
with a thermal velocity of $\sim 10^3$~m/s.
It turns out that in this case the accretion of the earth
by a mini black hole would take many times longer
than the age of the universe.
Thus, we have shown that for each of the growth estimates (\ref{Mgrowth},\ref{myvel})
the existence of white dwarfs is in contradiction with
the possibility of dangerous mini black holes.

\subsection*{{\bf{D2}}: higher emission energy}
There are two conditions to be fulfilled before arriving at this 
scenario. Which are
$(E_{ac}/t_{ac}>E_{em}/t_{em}=E_{1em}/t_{1em})$ and
 $(E_{1em}>m_e)$,
 where all magnitudes are defined in the rest frame of
 a mini black holes originating from a cosmic ray event and
 ($E_{1em}, t_{1em}$) are the average values per single emission.
Combining the two conditions
and solving for the emission time ($t_{1em}$) gives
\be\label{d4}
t_{1em}>t_{ac}\frac{m_e}{E_{ac}}\quad.
\ee
We will now show that this condition can not
be fulfilled by a self neutralizing mini black hole without
leading to electromagnetic stopping.
The minimal accretion energy for a high energetic 
nucleon black hole collision can be calculated like in equation
(\ref{Mgrowth2}) from the product of the proton radius, the minimal black
hole area (which is at $M=M_f$), and the proton density. 
This gives for one extra dimension ($E_{ac}(M=M_f,d=1)\approx 1.6\;10^{-3}$~GeV) and
it is slightly bigger for ($d>1$) until for six extra dimensions it is
($E_{ac}(M=M_f,d=6)\approx 4.7\;10^{-3}$~GeV).
Here one can read off a lower bound for the
accreted energy per event ($E_{ac}> 1.6\;10^{-3}$~GeV).
With this lower bound one finds
\be
t_{1em}>0.3\, t_{ac}\quad,
\ee
which shows that the emission time has to be at least
of the same order of magnitude as the accretion time.
The accretion time scale $t_{ac}$ is determined from the
mean distance $l_p$ (Lorentz contracted) which a mini black hole has
to travel in the sun until it encounters a nucleon as
$t_{ac}= (l_p /c) (2 m_p/M_f)\approx 10^{-22}$~s.
However, as discussed in ({\bf{D1-A}}) an emission time scale
which is comparable or bigger than the accretion time scale means
that electromagnetic stopping has to take place.
Therefore, the condition (\ref{d4}) inevitable leads to electromagnetic
stopping of mini black holes from cosmic rays which rules out
any danger from this scenario.

\subsection*{{\bf{D3}}: low emission energy}
In this discussion it is assumed that the mini
black hole emits less energy per emission than
the electron mass $E_{1em}<m_e\approx 511$~keV.
Since the electron is the lightest charged particle,
there is no way, the black hole can neutralize, once
it has obtained some charge due to its production or due
to a subsequent accretion.
Therefore, the above condition inevitable leads to electromagnetic
stopping of mini black holes from cosmic rays 
(like discussed in {\bf{D1-A}}) which rules out
any danger from this scenario.

\section{Summary}

In this paper we reviewed the framework for the
conjectured production of mini black holes at the LHC and
we have motivated the necessity of analyzing the possible
danger that could come with the production of mini black holes.
After this we discussed
 the (logically) possible black hole evolution paths.
Then we discussed every single outcome of those
paths (D0-D3) and showed that none of them 
can lead to a black hole disaster at the LHC.

\section*{Acknowledgements}
  This work was supported by GSI.
 The authors thank Michelangelo Mangano for fruitful discussions and his helpful comments.
   We also acknowledge discussion with Rainer Plaga.



\begin{thebibliography}{99}


\bibitem{ArkaniHamed:1998rs}
  N.~Arkani-Hamed, S.~Dimopoulos and G.~R.~Dvali,
  Phys.\ Lett.\  B {\bf 429}, 263 (1998)
  [arXiv:hep-ph/9803315].

\bibitem{Antoniadis:1998ig}
  I.~Antoniadis, N.~Arkani-Hamed, S.~Dimopoulos and G.~R.~Dvali,
  Phys.\ Lett.\  B {\bf 436}, 257 (1998)
  [arXiv:hep-ph/9804398].

\bibitem{ArkaniHamed:1998nn}
  N.~Arkani-Hamed, S.~Dimopoulos and G.~R.~Dvali,
  Phys.\ Rev.\  D {\bf 59}, 086004 (1999)
  [arXiv:hep-ph/9807344].

\bibitem{Randall:1999ee}
  L.~Randall and R.~Sundrum,
  Phys.\ Rev.\ Lett.\  {\bf 83}, 3370 (1999)
  [arXiv:hep-ph/9905221].

\bibitem{Randall:1999vf}
  L.~Randall and R.~Sundrum,
  Phys.\ Rev.\ Lett.\  {\bf 83}, 4690 (1999)
  [arXiv:hep-th/9906064].

\bibitem{Banks:1999gd}
  T.~Banks and W.~Fischler,
  arXiv:hep-th/9906038.

\bibitem{Giddings:2000ay}
  S.~B.~Giddings and E.~Katz,
  J.\ Math.\ Phys.\  {\bf 42}, 3082 (2001)
  [arXiv:hep-th/0009176].

\bibitem{Giddings:2001bu}
  S.~B.~Giddings and S.~D.~Thomas,
  Phys.\ Rev.\  D {\bf 65}, 056010 (2002)
  [arXiv:hep-ph/0106219].

\bibitem{Dimopoulos:2001hw}
  S.~Dimopoulos and G.~L.~Landsberg,
  Phys.\ Rev.\ Lett.\  {\bf 87}, 161602 (2001)
  [arXiv:hep-ph/0106295].

\bibitem{Hossenfelder:2001dn}
  S.~Hossenfelder, S.~Hofmann, M.~Bleicher and H.~Stoecker,
  Phys.\ Rev.\  D {\bf 66}, 101502 (2002)
  [arXiv:hep-ph/0109085].

\bibitem{Bleicher:2001kh}
  M.~Bleicher, S.~Hofmann, S.~Hossenfelder and H.~Stoecker,
  Phys.\ Lett.\  B {\bf 548}, 73 (2002)
  [arXiv:hep-ph/0112186].

\bibitem{Kotwal:2002wg}
  A.~V.~Kotwal and C.~Hays,
  Phys.\ Rev.\  D {\bf 66}, 116005 (2002)
  [arXiv:hep-ph/0206055].

\bibitem{Giddings:2004xy}
  S.~B.~Giddings and V.~S.~Rychkov,
  Phys.\ Rev.\  D {\bf 70}, 104026 (2004)
  [arXiv:hep-th/0409131].

\bibitem{blogs}
[http://backreaction.blogspot.com/2008/04/black-holes-at-lhc-again.html];
[http://www.youtube.com/watch?v=z7yZ5LEL5es];
[http://www.youtube.com/watch?v=ozjq80IF9dg];
[http://www.youtube.com/watch?v=BXzugu39pKM].


\bibitem{Giddings:2008gr}
  S.~B.~Giddings and M.~L.~Mangano,
  Phys.\ Rev.\  D {\bf 78}, 035009 (2008)
  [arXiv:0806.3381 [hep-ph]];
  S.~B.~Giddings and M.~L.~Mangano,
  arXiv:0808.4087 [hep-ph].


\bibitem{Bachacou:1999zb}
  H.~Bachacou, I.~Hinchliffe and F.~E.~Paige,
  Phys.\ Rev.\  D {\bf 62}, 015009 (2000)
  [arXiv:hep-ph/9907518].

\bibitem{Khoze:2001xm}
  V.~A.~Khoze, A.~D.~Martin and M.~G.~Ryskin,
  Eur.\ Phys.\ J.\  C {\bf 23}, 311 (2002)
  [arXiv:hep-ph/0111078].

\bibitem{Dimopoulos:2001qe}
  S.~Dimopoulos and R.~Emparan,
  Phys.\ Lett.\  B {\bf 526}, 393 (2002)
  [arXiv:hep-ph/0108060].

\bibitem{Weiglein:2004hn}
  G.~Weiglein {\it et al.}  [LHC/LC Study Group],
  Phys.\ Rept.\  {\bf 426}, 47 (2006)
  [arXiv:hep-ph/0410364].

\bibitem{ArkaniHamed:2004fb}
  N.~Arkani-Hamed and S.~Dimopoulos,
  JHEP {\bf 0506}, 073 (2005)
  [arXiv:hep-th/0405159].

\bibitem{Lillie:2007yh}
  B.~Lillie, L.~Randall and L.~T.~Wang,
  JHEP {\bf 0709}, 074 (2007)
  [arXiv:hep-ph/0701166].

\bibitem{Cheung:2007zza}
  K.~Cheung, W.~Y.~Keung and T.~C.~Yuan,
  Phys.\ Rev.\ Lett.\  {\bf 99}, 051803 (2007)
  [arXiv:0704.2588 [hep-ph]].

\bibitem{hoop} K.~S.~ Thorne,
 in Klauder, J., ed., Magic without Magic, 231-258,
(W. H. Freeman, San Francisco, 1972).

\bibitem{Myers:1986un}
  R.~C.~Myers and M.~J.~Perry,
  Annals Phys.\  {\bf 172}, 304 (1986).

\bibitem{Voloshin:2001fe}
  M.~B.~Voloshin,
  Phys.\ Lett.\  B {\bf 524}, 376 (2002)
  [Erratum-ibid.\  B {\bf 605}, 426 (2005)]
  [arXiv:hep-ph/0111099].

\bibitem{Voloshin:2001vs}
  M.~B.~Voloshin,
  Phys.\ Lett.\  B {\bf 518}, 137 (2001)
  [arXiv:hep-ph/0107119].

\bibitem{Giddings:2001ih}
  S.~B.~Giddings,
in {\it Proc. of the APS/DPF/DPB Summer Study on the Future of Particle Physics (Snowmass 2001) } ed. N.~Graf,
{\it In the Proceedings of APS / DPF / DPB Summer Study on the Future of Particle Physics (Snowmass 2001), Snowmass, Colorado, 30 Jun - 21 Jul
2001, pp P328}
  [arXiv:hep-ph/0110127].

\bibitem{Rizzo:2001dk}
  T.~G.~Rizzo,
in {\it Proc. of the APS/DPF/DPB Summer Study on the Future of Particle Physics (Snowmass 2001) } ed. N.~Graf,
{\it In the Proceedings of APS / DPF / DPB Summer Study on the Future of Particle Physics (Snowmass 2001), Snowmass, Colorado, 30 Jun - 21 Jul
2001, pp P339}
  [arXiv:hep-ph/0111230].

\bibitem{Jevicki:2002fq}
  A.~Jevicki and J.~Thaler,
  Phys.\ Rev.\  D {\bf 66}, 024041 (2002)
  [arXiv:hep-th/0203172].

\bibitem{Eardley:2002re}
  D.~M.~Eardley and S.~B.~Giddings,
  Phys.\ Rev.\  D {\bf 66}, 044011 (2002)
  [arXiv:gr-qc/0201034].

\bibitem{Rychkov:2004sf}
  V.~S.~Rychkov,
  Phys.\ Rev.\  D {\bf 70}, 044003 (2004)
  [arXiv:hep-ph/0401116].

\bibitem{Rychkov:2004tw}
  V.~S.~Rychkov,
  arXiv:hep-th/0410295.

\bibitem{Kang:2004yk}
  K.~Kang and H.~Nastase,
  Phys.\ Rev.\  D {\bf 71}, 124035 (2005)
  [arXiv:hep-th/0409099].

\bibitem{Rizzo:2006uz}
  T.~G.~Rizzo,
  Class.\ Quant.\ Grav.\  {\bf 23}, 4263 (2006)
  [arXiv:hep-ph/0601029].

\bibitem{Rizzo:2006zb}
  T.~G.~Rizzo,
  JHEP {\bf 0609}, 021 (2006)
  [arXiv:hep-ph/0606051].

\bibitem{Vachaspati:2006ki}
  T.~Vachaspati, D.~Stojkovic and L.~M.~Krauss,
  Phys.\ Rev.\  D {\bf 76}, 024005 (2007)
  [arXiv:gr-qc/0609024].

\bibitem{Vachaspati:2007hr}
  T.~Vachaspati and D.~Stojkovic,
  Phys.\ Lett.\  B {\bf 663}, 107 (2008)
  [arXiv:gr-qc/0701096].

\bibitem{Yoshino:2002tx}
  H.~Yoshino and Y.~Nambu,
  Phys.\ Rev.\  D {\bf 67}, 024009 (2003)
  [arXiv:gr-qc/0209003].

\bibitem{Solodukhin:2002ui}
  S.~N.~Solodukhin,
  Phys.\ Lett.\  B {\bf 533}, 153 (2002)
  [arXiv:hep-ph/0201248].

\bibitem{Ida:2002ez}
  D.~Ida, K.~y.~Oda and S.~C.~Park,
  Phys.\ Rev.\  D {\bf 67}, 064025 (2003)
  [Erratum-ibid.\  D {\bf 69}, 049901 (2004)]
  [arXiv:hep-th/0212108].

\bibitem{Horowitz:2002mw}
  G.~T.~Horowitz and J.~Polchinski,
  Phys.\ Rev.\  D {\bf 66}, 103512 (2002)
  [arXiv:hep-th/0206228].

\bibitem{Dar:1999ac}
  A.~Dar, A.~De Rujula and U.~W.~Heinz,
  Phys.\ Lett.\  B {\bf 470}, 142 (1999)
  [arXiv:hep-ph/9910471].

\bibitem{Whitt:1985ki}
  B.~Whitt,
  Phys.\ Rev.\  D {\bf 32}, 379 (1985).

\bibitem{Aharonov:1987tp}
  Y.~Aharonov, A.~Casher and S.~Nussinov,
  Phys.\ Lett.\  B {\bf 191}, 51 (1987).

\bibitem{Gibbons:1987ps}
  G.~W.~Gibbons and K.~i.~Maeda,
  Nucl.\ Phys.\  B {\bf 298}, 741 (1988).

\bibitem{Whitt:1988ax}
  B.~Whitt,
  Phys.\ Rev.\  D {\bf 38}, 3000 (1988).

\bibitem{Bowick:1988xh}
  M.~J.~Bowick, S.~B.~Giddings, J.~A.~Harvey, G.~T.~Horowitz and A.~Strominger,
  Phys.\ Rev.\ Lett.\  {\bf 61}, 2823 (1988).

\bibitem{Callan:1988hs}
  C.~G.~.~Callan, R.~C.~Myers and M.~J.~Perry,
  Nucl.\ Phys.\  B {\bf 311}, 673 (1989).

\bibitem{Myers:1988ze}
  R.~C.~Myers and J.~Z.~Simon,
  Phys.\ Rev.\  D {\bf 38}, 2434 (1988).

\bibitem{Coleman:1991sj}
  S.~R.~Coleman, J.~Preskill and F.~Wilczek,
  Mod.\ Phys.\ Lett.\  A {\bf 6}, 1631 (1991).

\bibitem{Lee:1991qs}
  K.~M.~Lee, V.~P.~Nair and E.~J.~Weinberg,
  Phys.\ Rev.\ Lett.\  {\bf 68}, 1100 (1992)
  [arXiv:hep-th/9111045].

\bibitem{Banks:1992ba}
  T.~Banks, A.~Dabholkar, M.~R.~Douglas and M.~O'Loughlin,
  Phys.\ Rev.\  D {\bf 45}, 3607 (1992)
  [arXiv:hep-th/9201061].

\bibitem{Barrow:1992hq}
  J.~D.~Barrow, E.~J.~Copeland and A.~R.~Liddle,
  Phys.\ Rev.\  D {\bf 46}, 645 (1992).

\bibitem{Banks:1992is}
  T.~Banks, M.~O'Loughlin and A.~Strominger,
  Phys.\ Rev.\  D {\bf 47}, 4476 (1993)
  [arXiv:hep-th/9211030].

\bibitem{Maeda:1993ap}
  K.~I.~Maeda, T.~Tachizawa, T.~Torii and T.~Maki,
  Phys.\ Rev.\ Lett.\  {\bf 72}, 450 (1994)
  [arXiv:gr-qc/9310015].

\bibitem{Giddings:1993vj}
  S.~B.~Giddings,
  Phys.\ Rev.\  D {\bf 49}, 4078 (1994)
  [arXiv:hep-th/9310101].

\bibitem{Bonanno:2000ep}
  A.~Bonanno and M.~Reuter,
  Phys.\ Rev.\  D {\bf 62}, 043008 (2000)
  [arXiv:hep-th/0002196].

\bibitem{Adler:2001vs}
  R.~J.~Adler, P.~Chen and D.~I.~Santiago,
  Gen.\ Rel.\ Grav.\  {\bf 33}, 2101 (2001)
  [arXiv:gr-qc/0106080].

\bibitem{Alexeyev:2002tg}
  S.~Alexeyev, A.~Barrau, G.~Boudoul, O.~Khovanskaya and M.~Sazhin,
  Class.\ Quant.\ Grav.\  {\bf 19}, 4431 (2002)
  [arXiv:gr-qc/0201069].

\bibitem{Baker:2003ds}
  J.~G.~Baker, M.~Campanelli, C.~O.~Lousto and R.~Takahashi,
  Phys.\ Rev.\  D {\bf 69}, 027505 (2004)
  [arXiv:astro-ph/0305287].

\bibitem{Hossenfelder:2003dy}
  S.~Hossenfelder, M.~Bleicher, S.~Hofmann, H.~Stoecker and A.~V.~Kotwal,
  Phys.\ Lett.\  B {\bf 566}, 233 (2003)
  [arXiv:hep-ph/0302247].

\bibitem{Koch:2005ks}
  B.~Koch, M.~Bleicher and S.~Hossenfelder,
  JHEP {\bf 0510}, 053 (2005)
  [arXiv:hep-ph/0507138].

\bibitem{Hossenfelder:2005ku}
  S.~Hossenfelder, B.~Koch and M.~Bleicher,
  arXiv:hep-ph/0507140.

\bibitem{Humanic:2006xg}
  T.~J.~Humanic, B.~Koch and H.~Stoecker,
  Int.\ J.\ Mod.\ Phys.\  E {\bf 16}, 841 (2007)
  [arXiv:hep-ph/0607097].

\bibitem{Ghaffarnejad:2007tm}
  H.~Ghaffarnejad,
  Phys.\ Rev.\  D {\bf 75}, 084009 (2007).

\bibitem{Li:2007ga}
  X.~Li,
  Phys.\ Lett.\  B {\bf 647}, 207 (2007).

\bibitem{Koch:2007yt}
  B.~Koch,
  arXiv:0707.4644 [hep-ph].

\bibitem{Nicolini:2008aj}
  P.~Nicolini,
  arXiv:0807.1939 [hep-th].

\bibitem{Arkani:2008}
  N.~Arkani Hamed,
  ``Asking a Judge to Save the World, and Maybe a Whole Lot More,''
  New York Times, {\bf sciene}, March 29 (1999).

\bibitem{Ringwald:2001vk}
  A.~Ringwald and H.~Tu,
  Phys.\ Lett.\  B {\bf 525}, 135 (2002)
  [arXiv:hep-ph/0111042].

\bibitem{Hawking:1974sw}
  S.~W.~Hawking,
  Commun.\ Math.\ Phys.\  {\bf 43}, 199 (1975)
  [Erratum-ibid.\  {\bf 46}, 206 (1976)].

\bibitem{Hawking:1982dh}
  S.~W.~Hawking and D.~N.~Page,
  Commun.\ Math.\ Phys.\  {\bf 87}, 577 (1983).


\bibitem{plaga}
 R.~Plaga,
Private communication,
(2008).

\bibitem{Plaga:2008gq}
  R.~Plaga,
  arXiv:0808.1415 [hep-ph].

\end{thebibliography}
\end{document}